# Ligand Binding, Protein Fluctuations, And Allosteric Free Energy

Michael E. Wall

*Computer and Computational Sciences Division & Bioscience Division, Los Alamos National Laboratory, Los Alamos, NM 87545 USA. E-mail: mewall@lanl.gov*

**Abstract.** Although the importance of protein dynamics in protein function is generally recognized, the role of protein fluctuations in allosteric effects scarcely has been considered. To address this gap, the Kullback-Leibler divergence ($D_\mathbf{x}$) between protein conformational distributions before and after ligand binding was proposed as a means of quantifying allosteric effects in proteins. Here, previous applications of $D_\mathbf{x}$ to methods for analysis and simulation of proteins are first reviewed, and their implications for understanding aspects of protein function and protein evolution are discussed. Next, equations for $D_\mathbf{x}$ suggest that $k_B T D_\mathbf{x}$ should be interpreted as an allosteric free energy – the free energy associated with changing the ligand-free protein conformational distribution to the ligand-bound conformational distribution. This interpretation leads to a thermodynamic model of allosteric transitions that unifies existing perspectives on the relation between ligand binding and changes in protein conformational distributions. The definition of $D_\mathbf{x}$ is used to explore some interesting mathematical relations among commonly recognized thermodynamic and biophysical quantities, such as the total free energy change upon ligand binding, and ligand-binding affinities for individual protein conformations. These results represent the beginnings of a theoretical framework for considering the full protein conformational distribution in modeling allosteric transitions. Early applications of the framework have produced results with implications both for methods for coarsed-grained modeling of proteins, and for understanding the relation between ligand binding and protein dynamics.

## INTRODUCTION

One important mechanism of protein regulation is allosteric regulation, in which molecular interactions influence protein activity through changes in protein structure. In traditional models of allosteric regulation, proteins adopt a limited number of conformations, each of which may have a different activity [1,2]. However, the importance of considering continuous conformational distributions in understanding allosteric effects was recognized by Weber [3]. Neutron scattering experiments later provided evidence for changes in protein dynamics upon ligand binding [4], and it was subsequently realized that ligand binding at an allosteric site can influence binding at a remote site without inducing a mean conformational change, solely through alteration of atomic fluctuations [5]. Indeed, the conformational distribution is known to be a key determinant of protein activity [6], and is a key element in rate theories [7].

Motivated by these considerations, a theoretical framework was recently developed to quantify changes in protein conformational distributions upon ligand binding in terms of the Kullback-Leibler divergence [8], $D_\mathbf{x}$ [9,10]. A closed-form estimate of $D_\mathbf{x}$ was derived in the harmonic approximation [10], and was used to demonstrate that values of $D_\mathbf{x}$ are elevated in small-molecule binding sites of proteins [11]. To analyze aspects of allosteric mechanisms, $D_\mathbf{x}$ was calculated for local regions of bovine trypsinogen upon binding ligands in the active site or allosteric site of the protein. In addition, $D_\mathbf{x}$ has been used to develop rigorous methods for evaluating and optimizing coarse-grained models of proteins [10]. These previous results are reviewed below.

While previous work has focused on the application of $D_\mathbf{x}$ to methods for analysis and simulation of proteins, the biophysical meaning of $D_\mathbf{x}$ has not yet been described in detail. Therefore, after the review of previous results is a discussion of aspects of the thermodynamic and biophysical significance of $D_\mathbf{x}$, with implications for understanding allosteric effects in proteins.

## DEFINITION OF $D_\mathbf{X}$

The Kullback-Leibler divergence $D_\mathbf{x}$ upon binding a ligand is defined as

$$D_\mathbf{x} = \int d^{3N}\mathbf{x}\ P'(\mathbf{x})\ln\frac{P'(\mathbf{x})}{P(\mathbf{x})} \tag{1}$$

where $P'(\mathbf{x})$ is the protein conformational distribution in the presence of the ligand, and $P(\mathbf{x})$ is the same distribution in the absence of the ligand [9]. It is easily shown that $D_\mathbf{x}$ is always non-negative. Note that $D_\mathbf{x}$ is not a distance measure because its value is not conserved when $P'(\mathbf{x})$ and $P(\mathbf{x})$ are interchanged in Eq. (1).

Let a reaction rate $k(\mathbf{x})$ be a function of the configuration $\mathbf{x}$ of $N$ atoms. The relation between the rate distribution $P(k)$ and the conformational distribution $P(\mathbf{x})$ is then [9]

$$P(k) = \int d^{3N}\mathbf{x} P(\mathbf{x})\delta[k(\mathbf{x}) - k], \tag{2}$$

where the integral is over all conformations $\mathbf{x}$. Equation (2) concisely expresses the necessity of considering the full conformational distribution in understanding the functional consequences of allosteric transitions.

## $D_\mathbf{X}$ IN THE HARMONIC APPROXIMATION

To calculate $D_\mathbf{x}$ using Eq. (1), it is necessary to calculate the marginal probability distribution of the protein configurations $P(\mathbf{x})$ from a full conformational distribution $P(\mathbf{x},\mathbf{y})$ of a protein-ligand complex,

$$P(\mathbf{x}) = \int d^{3N_y}\mathbf{y} P(\mathbf{x},\mathbf{y}). \tag{3}$$

In Eq. (3), $\mathbf{y}$ is a vector of the $N_y$ ligand coordinates. Solutions for Equations (3) and (1) have been derived in a model of harmonic vibrations of the protein-ligand system [9,10]. Let $\mathbf{z} = (\mathbf{x},\mathbf{y})$ be the coordinates of the combined protein-ligand system, measured relative to an equilibrium configuration $\mathbf{z}_0 = (\mathbf{x}_0, \mathbf{y}_0)$. Consider a harmonic approximation to the potential energy function $U(\mathbf{z}_0 + \mathbf{z})$,

$$U(\mathbf{z}_0 + \mathbf{z}) \approx U(\mathbf{z}_0) + \frac{1}{2}\mathbf{z}^T\mathbf{Hz}, \qquad (4)$$

where $\mathbf{H}$ is the Hessian of $U$ evaluated at $\mathbf{z}_0$: $H_{ij}\big|_{\mathbf{z}_0} = \partial^2 U/\partial z_i \partial z_j \big|_{\mathbf{z}_0}$. Assuming a Boltzmann distribution for $P(\mathbf{z})$ and ignoring solvent and pressure effects,

$$\begin{aligned} P(\mathbf{z}) &= Z^{-1} e^{-\mathbf{z}^T \mathbf{Hz}/2k_B T} \\ &= (2\pi k_B T)^{-3N_z/2} e^{-\left|\Omega \mathbf{V}^T \mathbf{z}\right|^2 / 2k_B T} \prod_{\substack{i=1\ldots 3N_z \\ \omega_i \neq 0}} \omega_i, \end{aligned} \qquad (5)$$

where $Z$ is the partition function, $k_B$ is Boltzmann's constant, $T$ is the temperature, $N_z$ is the number of atoms in the complex, the elements of the matrix $\Omega^2 = diag(\omega_1^2,\ldots,\omega_{3N_z}^2)$ are the eigenvalues of $\mathbf{H}$, and the columns of the matrix $\mathbf{V}$ are the eigenvectors of $\mathbf{H}$. Here and elsewhere, products and summations are carried out over nonzero modes. Define the submatrices $\mathbf{H_x}$, $\mathbf{H_y}$, and $\mathbf{G}$ as

$$\mathbf{Hz} = \begin{pmatrix} \mathbf{H_x} & \mathbf{G} \\ \mathbf{G}^T & \mathbf{H_y} \end{pmatrix} \begin{pmatrix} \mathbf{x} \\ \mathbf{y} \end{pmatrix} = \begin{pmatrix} \mathbf{H_x x} + \mathbf{Gy} \\ \mathbf{G}^T \mathbf{x} + \mathbf{H_y y} \end{pmatrix}. \qquad (6)$$

$\mathbf{H_x}$ couples coordinates from $\mathbf{x}$, $\mathbf{H_y}$ couples coordinates from $\mathbf{y}$, and $\mathbf{G}$ couples coordinates between $\mathbf{x}$ and $\mathbf{y}$. By re-expressing $\left|\Omega\mathbf{V}^T\mathbf{z}\right|^2$ in Eq. (5) using Eq. (6), the integral in Eq. (3) may be performed, yielding

$$P(\mathbf{x}) = (2\pi k_B T)^{-3N/2} e^{-\left|\overline{\Omega}\overline{\mathbf{V}}^T \mathbf{z}\right|^2 / 2k_B T} \prod_{\substack{i=1\ldots 3N \\ \overline{\omega}_i \neq 0}} \overline{\omega}_i. \qquad (7)$$

In Eq. (7), the elements of the matrix $\overline{\Omega}^2 = diag(\overline{\omega}_1^2,\ldots,\overline{\omega}_{3N}^2)$ and the columns $\overline{\mathbf{v}}_i$ of the matrix $\overline{\mathbf{V}}$ are the eigenvalues and eigenvectors of a matrix $\overline{\mathbf{H}}$, defined as

$$\overline{\mathbf{H}} = \mathbf{H_x} - \mathbf{G H_y^{-1} G}^T = \overline{\mathbf{V}}\left|\overline{\Omega}\right|^2 \overline{\mathbf{V}}^T. \qquad (8)$$

Equation (8) was independently derived by Ming & Wall [10] and by Zheng & Brooks [12]. When there is no interaction between the protein and ligand coordinates, Eq. (7) is just the conformational distribution of the protein in the absence of the ligand; let the unprimed probability distribution $P(\mathbf{x})$ in Eq. (7) correspond to this case. When there is an interaction between the protein and the ligand, let the distribution $P'(\mathbf{x})$ be given by a similar expression,

$$P'(\mathbf{x}) = (2\pi k_B T)^{-3N/2} e^{-\left|\overline{\Omega}'\overline{\mathbf{V}}'^T \mathbf{z}\right|^2 / 2k_B T} \prod_{\substack{i=1\ldots 3N \\ \overline{\omega}'_i \neq 0}} \overline{\omega}'_i. \qquad (9)$$

With respect to the unprimed variables in Eq. (7), the primed variables in Eq. (9) correspond to the case in which the protein and ligand interact. Using methods described in detail in Ref. [9], substituting Eqs. (7) and (9) in Eq. (1) yields the following expression for $D_\mathbf{x}$:

$$D_{\mathbf{x}} = \sum_{i=1...3N}^{\overline{\omega}_i' \neq 0; \overline{\omega}_i \neq 0} \left( \log \frac{\overline{\omega}_i'}{\overline{\omega}_i} + \frac{1}{2k_B T} \overline{\omega}_i^2 |\Delta \mathbf{x}_0 \cdot \overline{\mathbf{v}}_i|^2 + \frac{1}{2} \sum_{j=1...3N}^{\overline{\omega}_j \neq 0} \frac{\overline{\omega}_j^2}{\overline{\omega}_i'^2} |\overline{\mathbf{v}}_i' \cdot \overline{\mathbf{v}}_j|^2 - \frac{1}{2} \right). \quad (10)$$

The first term is proportional to the entropy change upon releasing the ligand, and the second term is proportional to the potential energy required to change the mean conformation of the protein without the ligand to that with the ligand. Equation (10) and variants thereof have been used in several practical applications [10,11], which are reviewed in the next section.

## PRACTICAL APPLICATIONS OF $D_X$

### Prediction Of Ligand-Binding Sites

Recently Ming & Wall [11] examined 305 protein structures from the GOLD docking test set [13] and investigated whether interactions at small-molecule binding sites cause a large change in the protein conformational distribution. A computational method, called dynamics perturbation analysis (DPA), was presented to identify sites at which interactions yield a large value of $D_\mathbf{x}$. DPA was used to analyze proteins in the test set, and to determine whether $D_\mathbf{x}$ values for points in the neighborhood of ligand-binding sites were high compared to random points. A method was then developed to predict functional sites in proteins, and the method was evaluated using proteins in the test set. The performance of the method was compared to that of a cleft analysis method.

The DPA method was based on a method previously used to analyze changes in molecular vibrations of a lysozyme-NAG complex for random protein-ligand interactions [9]. In DPA, a protein is decorated with $M$ surface points that interact with neighboring protein atoms. The protein conformational distribution $P^{(0)}(\mathbf{x})$ is calculated in the absence of any surface points, and $M$ protein conformational distributions $P^{(m)}(\mathbf{x})$ are calculated for the protein interacting with each point $m$. The conformational distributions are calculated using a coarse-grained model of molecular vibrations, and the distributions $P^{(m)}(\mathbf{x})$ are calculated from models of the protein in complex with each surface point using Eq. (9). The Kullback-Leibler divergence $D_\mathbf{x}^{(m)}$ between $P^{(0)}(\mathbf{x})$ and $P^{(m)}(\mathbf{x})$ is calculated for each point $m$ (Eq. (10)), and is used as a measure of the change in the protein conformational distribution upon interacting with point $m$.

Protein vibrations were modeled using the elastic network model (ENM) [14-17]. In the ENM, alpha-carbon atoms are extracted from an atomic model of a protein, and an interaction network is generated by connecting springs between all atom pairs separated by a distance less than or equal to a cutoff distance $r_c$. Each spring has the same force constant $\gamma$, is aligned with the separation between the connected atoms, and has an equilibrium length equal to the equilibrium distance between the atoms. The interaction between the protein and a surface point $m$ was modeled by connecting springs of force constant $\gamma_s$ between the surface point and all protein atoms within a cutoff distance $r_s$ of the surface point. The protein coordinates were not modified in modeling the interaction.

Consistent with results obtained using an all-atom model of lysozyme [9], values of $D_x^{(m)}$ were found to be elevated in the neighborhood of the tri-N-acetyl-D-glucosamine (tri-NAG) binding site. Interestingly, the distribution of $y = D_x^{(m)}$ values was empirically well-fit by a probability density ρ(y) given by

$$\rho(y) = \frac{1}{\beta} e^{\frac{y-\mu}{\beta} - e^{\frac{y-\mu}{\beta}}}. \qquad (11)$$

which is an extreme value distribution of width β centered on μ (Fig. 1).

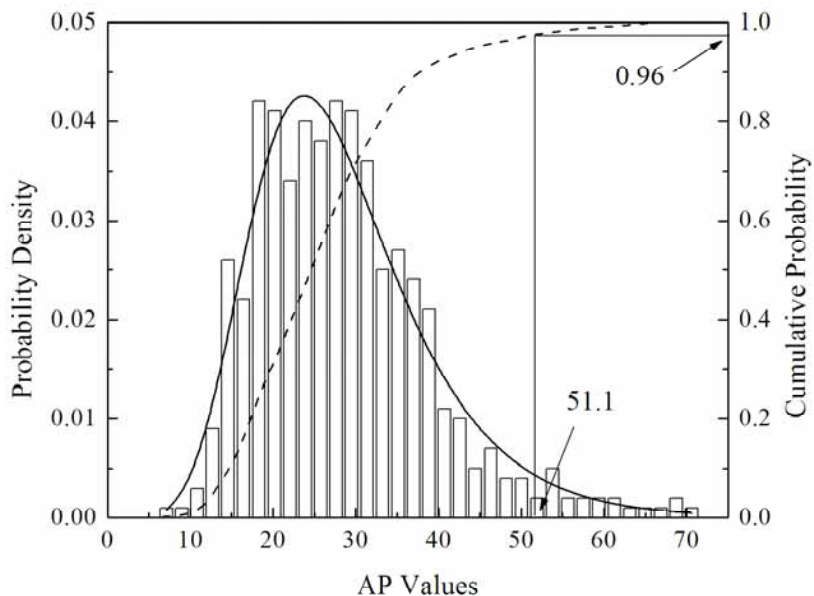

**FIGURE 1.** Distribution of $D_x^{(m)}$ values (labeled as AP values) for 4859 points on the surface of lysozyme (the number of points was increased in this case to evaluate the fit). The distribution is well-fit by an extreme value distribution with parameters μ = 23.07 and β = 8.45 (Pearson correlation coefficient of 0.992). The fit is used to find the 96% upper bound of $D_x^{(m)}$ for the surface points; this bound is used as the threshold to select high-$D_x^{(m)}$ points for use in predicting functional sites.

DPA was applied to 305 protein structures in the GOLD docking test set [13]. Calculations were performed in the same manner as for lysozyme. A statistical analysis method was developed to quantitatively assess the tendency for $D_x^{(m)}$ values to be elevated in the neighborhood of ligand-binding sites. In this method, surface points in the neighborhood of the ligand were selected, each point was ranked with respect to all other surface points in terms of the value of $D_x^{(m)}$, and a composite score for all points was calculated. The probability of obtaining the composite score by randomly selecting points on the surface was calculated and was used to calculate a P-value for evaluating the significance of the score for each protein. In 14 of the 305 proteins, the ligand was buried and was not close to any of the surface points. Results for the rest of the proteins are illustrated in Fig. 2. For 95% of proteins, the P-value is $10^{-3}$ or lower, indicating that the elevation of $D_\mathbf{x}^{(m)}$ in the neighborhood of functional sites was statistically significant.

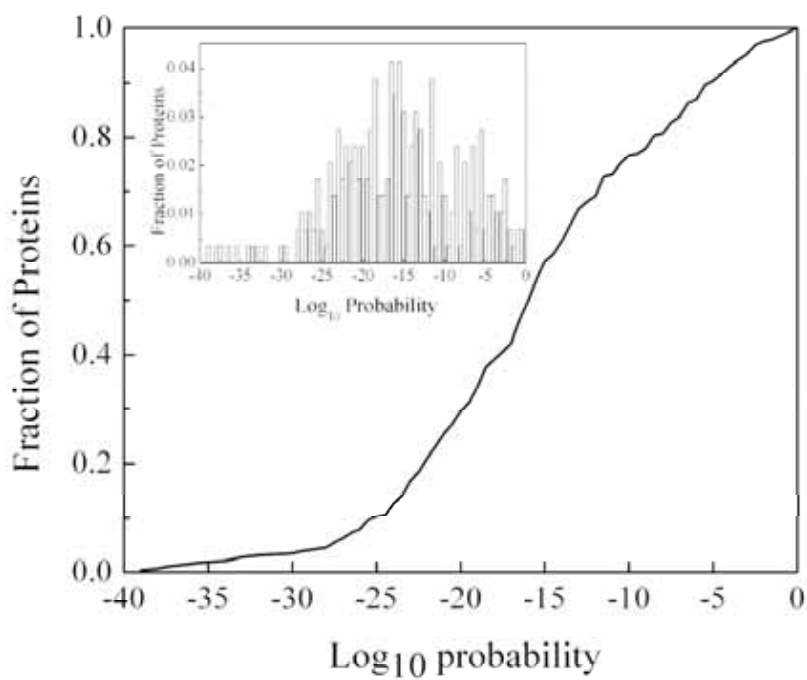

**FIGURE 2**. Statistical significance of elevated values of $D_\mathbf{x}^{(m)}$ in functional sites. The distribution of P-values (calculated in bins of width 2 in log units) calculated for the composite $D_\mathbf{x}^{(m)}$ score is shown for 291 proteins in the GOLD docking test set. For 95% of proteins, the P-value is $10^{-3}$ or lower, indicating that the elevation of $D_\mathbf{x}^{(m)}$ in the neighborhood of functional sites is statistically significant.

Following the weaker suggestion from all-atom study of lysozyme [9], these results strongly suggested that points with high values of $D_\mathbf{x}^{(m)}$ could be used to predict the locations of functional sites, and motivated the development of an algorithm for this purpose. The algorithm works as follows. First, DPA is performed on a protein. Then, the statistics of $D_\mathbf{x}^{(m)}$ values is modeled using an extreme value distribution. Points with significantly high values of $D_\mathbf{x}^{(m)}$ are selected and are spatially clustered. The clusters are ranked according to the mean value of $D_\mathbf{x}^{(m)}$ within the cluster, and points in the highest-ranked cluster are predicted to be associated with a functional site. Finally, residues in the neighborhood of the highest-ranked cluster are selected and are predicted to reside within the functional site.

Consistent with the analysis of lysozyme, the $D_\mathbf{x}^{(m)}$ values for the test-set proteins indicated that the statistics are well-described by an extreme-value distribution. To select points with significantly high values of $y = D_\mathbf{x}^{(m)}$, an operating point $C$ of the cumulative distribution $c(y) = 1 - e^{-e^{\frac{y-\mu}{\beta}}}$ was selected, $\mu$ and $\beta$ were fitted using the actual distribution of $D_\mathbf{x}^{(m)}$ for the protein, and a lower threshold $Y$ on $D_\mathbf{x}^{(m)}$ was calculated. Points with $D_\mathbf{x}^{(m)} > Y$ were clustered spatially, and the mean value of $D_\mathbf{x}^{(m)}$ for each cluster was calculated and was used to rank the clusters; for each protein, the rank-1 cluster was identified as the cluster with the highest mean value.

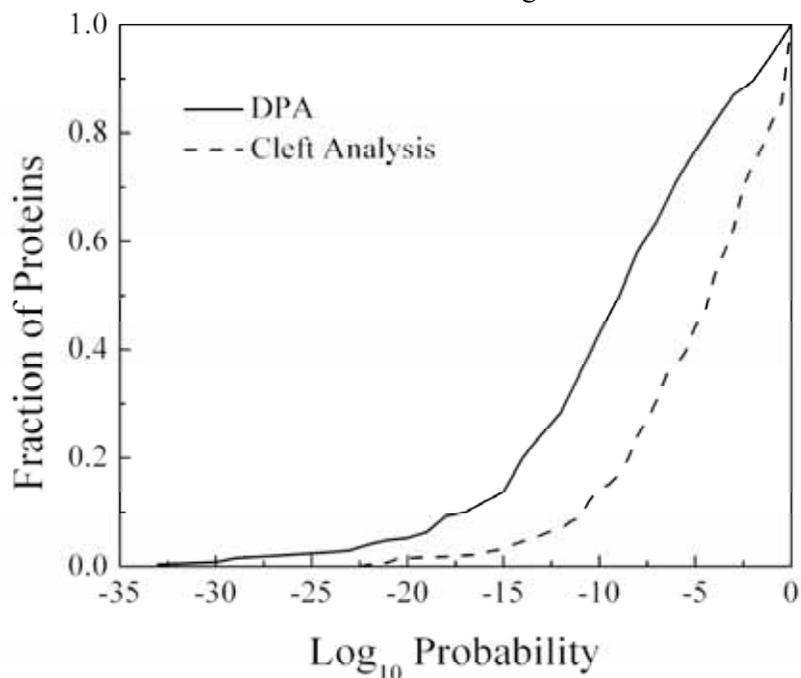

**FIGURE 3**. Statistical significance of the overlaps of predicted residues with ligand-binding-site residues. For each protein, a P-value (corresponding to the probability in a null model of finding at least as many ligand-binding-site resides as does the prediction algorithm) is calculated; the resulting distribution of P-values is shown here. For the DPA algorithm (solid line), a total of 250 proteins in the test set were considered; and for the cleft analysis algorithm (dashed line), a total of 278 proteins were considered. (In each case, only proteins for which the algorithm yielded at least one residue in the ligand-binding site were considered.)

Protein alpha-carbons within 6 Å of any of the points in the rank-1 cluster were selected and were used to identify the set of residues that are predicted to reside in a functional site. To evaluate the predictions, they were compared with the set of residues that are in the neighborhood of the ligand found in complex with the protein in the test set.

The method produced predictions for 287 of the 305 proteins. In 87% of cases (250 proteins), at least one predicted residue was in the ligand-binding site. The recall was at least 0.3 for 80% of cases, and was at least 0.5 for 76% of the cases. The precision was at least 0.3 for 68% of the cases, and was at least 0.5 for 44% of the cases. The statistical significance of the overlaps was assessed using a null model in which surface residues were randomly selected. Using the null model, a P-value was calculated to evaluate predictions for the 250 proteins in which at least one predicted residue was in the ligand-binding site. Results are shown in Fig. 3. For 87% of the cases, the P-value is $10^{-3}$ or smaller, indicating that there is a statistically significant overlap. As shown in Fig. 3, the performance of the DPA method compared favorably to that of a cleft analysis method for predicting ligand-binding residues.

## Optimization Of Coarse-Grained Molecular Models

In one common coarse-graining method, an all-atom model is simplified by considering effective interactions among a subset of the atoms (e.g., just the alpha-carbons). The usual criterion for model accuracy is the ability of a model to reproduce atomic mean-squared displacements (MSDs). However, MSDs are just one aspect of protein dynamics -- a stricter criterion for the accuracy of a coarse-grained model is the similarity between the configurational distributions of the selected atoms in the coarse-grained and all-atom models. Such a criterion is also biologically relevant, in part because the conformational distribution is a key determinant of protein activity [6,7,18] (Eq. (2)).

One useful measure of the difference between conformational distributions obtained from all-atom and coarse-grained simulations is the Kullback-Leibler divergence $D_\mathbf{x}$ [10]. Let $\mathbf{x}_\alpha$ be the coordinates of the $N_\alpha$ alpha-carbons in either an all-atom model of molecular vibrations, or a coarse-grained model with parameters $\Gamma$. Define the optimal coarse-grained model as the one for which $D_\mathbf{x}$ between $P^{(\Gamma)}(\mathbf{x}_\alpha)$ and $P(\mathbf{x}_\alpha)$ is minimal, i.e., for which $\Gamma$ is chosen such that

$$D_{\mathbf{x}_\alpha}^{(\Gamma)} = \sum_{i=1\ldots 3N_\alpha}^{\omega_i^{(\Gamma)} \neq 0; \overline{\omega}_i \neq 0} \left( \log \frac{\omega_i^{(\Gamma)}}{\overline{\omega}_i} + \frac{1}{2k_B T} \overline{\omega}_i^2 \left| \Delta \mathbf{x}_{\alpha,0} \cdot \overline{\mathbf{v}}_i \right|^2 \right. \\ \left. + \frac{1}{2} \sum_{j=1\ldots 3N_\alpha}^{\overline{\omega}_j \neq 0} \frac{\overline{\omega}_j^2}{\omega_i^{(\Gamma)2}} \left| \mathbf{v}_i^{(\Gamma)} \cdot \overline{\mathbf{v}}_j \right|^2 - \frac{1}{2} \right)$$

(12)

is minimal. In Eq. (12), $\omega_i^{(\Gamma)2}$ and $\mathbf{v}_i^{(\Gamma)}$ are the eigenvalue and eigenvector of mode *i* of the coarse-grained model; $\overline{\omega}_j^2$ and $\overline{\mathbf{v}}_j$ are the $i^{th}$ eigenvalue and eigenvector of the matrix $\overline{\mathbf{H}}$ calculated for the alpha-carbon atoms of the all-atom model (Eq. (8)), and $\Delta \mathbf{x}_{\alpha,0}$ is the difference between the equilibrium coordinates of the coarse-grained and all-atom models.

Ming & Wall [10] used Eq. (12) to find optimal solutions to the elastic network model (ENM) of protein vibrations [14,17], in which interacting alpha-carbons are connected by springs aligned with the direction of atomic separation. Following the Tirion model [17], each spring has the same force constant $\gamma$. For a given interaction network, the eigenvectors $\mathbf{v}_i^{(\Gamma)}$ are independent of $\gamma$, and each eigenvalue $\omega_i^{(\Gamma)2}$ is proportional to $\gamma$. Assuming no difference in the mean conformation between the models, the value of $\gamma$ at which $D_{\mathbf{x}_\alpha}^{(\Gamma)}$ is minimal was calculated as

$$\gamma = \frac{1}{3N_\alpha} \sum_{i=1\ldots 3N_\alpha}^{a_i \neq 0} \sum_{j=1\ldots 3N_\alpha}^{\overline{\omega}_j \neq 0} \frac{\overline{\omega}_j^2}{a_i^2} \left| \mathbf{v}_i^{(\Gamma)} \cdot \overline{\mathbf{v}}_j \right|^2,$$

(13)

where the constants $a_i^2 = \omega_i^{(\Gamma)2}/\gamma$ are independent of $\gamma$. Interestingly, the third and fourth terms of Eq. (12) cancel for this value of $\gamma$. Therefore, the optimal coarse-grained model is one that uses an energy scale for interactions that eliminates the contribution to $D_{\mathbf{x}_\alpha}^{(\Gamma)}$ due to differences in the eigenvectors, and which, given this energy scale, maximizes the entropy of the conformational distribution $P^{(\Gamma)}(\mathbf{x}_\alpha)$.

- 8 -

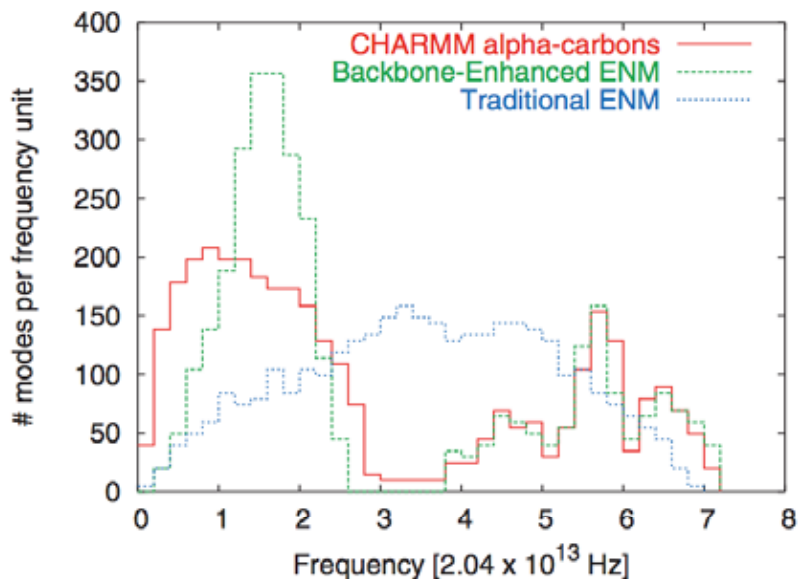

**FIGURE 4**. Density-of-states distribution for all-atom and elastic network models of trypsinogen. Frequency units are (Kcal / mol Å$^2$ $m_p$)$^{1/2}$ = 2.04 × 10$^{13}$ Hz, where $m_p$ is the proton mass. Densities were estimated by counting the number of modes in bins of width 0.2, and normalizing the integral to 663, which is the total number of non-zero modes. The ENM (dotted line) does not reproduce the bimodal distribution from the all-atom model (solid line); however, the BENM recovers the bimodal distribution (dashed line).

When the above method was applied to optimize an ENM of trypsinogen, the value of $D_{\mathbf{x}_\alpha}^{(\Gamma)} = 313$ was quite high [10]. This finding led to the discovery of a discrepancy between the density of states distributions from the ENM and the all-atom model: whereas the all-atom model yielded a bimodal distribution, the ENM yielded a unimodal distribution (Fig. 4). To correct this discrepancy, a backbone-enhanced elastic network model (BENM) was proposed, in which the spring constant was increased for alpha carbons that are backbone neighbors. The optimal BENM recovered the correct bimodal distribution of the density of states, and yielded a much lower value of $D_{\mathbf{x}_\alpha}^{(\Gamma)} = 102$. Because the agreement at high frequencies was especially good (Fig. 4), the remaining differences are most likely dominated by inaccuracies in the BENM in modeling low-frequency, large-scale vibrations.

## Analysis Of Allosteric Mechanisms

As demonstrated by the work of Hilser, Freire, and coworkers [19-22], it is interesting to analyze communication between remote sites in proteins in terms of conformational ensembles. In their framework, instead of describing the full atomic configurational distribution, the conformational ensemble is simplified to indicate whether residues are in a folded or unfolded state. This simplification, however, enables conformational ensembles to be calculated using methods that yield good agreement with experimentally observed hydrogen exchange protection factors for individual residues [23]. Using this framework, cooperativity effects in residue perturbations [19,22], ligand-binding [21], and intrinsic correlations in the conformational ensemble [20] have been characterized, yielding insight into allosteric effects in proteins.

Hawkins & McLeish recently have analyzed allosteric effects in terms of configurational entropy changes in Lac repressor [24]. In this study, changes in the entropy of low-frequency vibrations of the protein upon binding an allosteric regulator were calculated in a coarse-grained model. They have also analyzed vibrational free energies in coiled-coil structures such as dynein [25]. Filamentous protein structures were modeled using continuum elastic models of intertwined rods, enabling calculations of free energies due to thermal vibrations. Vibrational free energies were calculated in the presence and absence of clamping interactions at the ends of the rods, and were used to yield insight into allosteric effects.

Recently, Ming & Wall [10] used Eq. (10) to develop a general framework in which allosteric effects may be analyzed using the full configurational distribution in the harmonic approximation. In the spirit of Luque & Freire's analysis of IIAGlc binding to glycerol kinase [21], changes in the configurational distribution of local regions of trypsinogen were calculated upon binding Val-Val in an allosteric site, and bovine pancreatic trypsinogen inhibitor (BPTI) in the active site. The BENM coarse-grained model was used for calculations. In binding BPTI, it was found that the local values of $D_\mathbf{x}$ were relatively large in the neighborhood of the BPTI-binding site. Values of $D_\mathbf{x}$ elsewhere on the surface were smaller, with one interesting exception: values in the Val-Val binding site, which is an allosteric site, were comparable to those in the BPTI-binding site. In addition, binding of Val-Val in the allosteric site yielded a relatively large value of $D_\mathbf{x}$ in the neighborhood of Ser 195, which is the key catalytic residue for trypsin and other serine proteases: the value of $D_\mathbf{x}$ in this neighborhood was the 40$^{th}$ highest of 223 residues in the crystal structure; 11$^{th}$ of all residues not directly interacting with the Val-Val in the model; the highest of all residues located at least as far as Ser 195 is from the Val-Val ligand; and greater than that for 20 of 60 residues located closer to the ligand. These results indicated that there is a relatively strong communication between the regulatory and active sites of trypsinogen.

## THERMODYNAMIC INTERPRETATION OF $D_\mathbf{X}$

### Allosteric Free Energy

Let $G(x)$ be the Gibbs free energy of a protein in conformation $x$ in solution at constant temperature and pressure. (Here we use $x$ as a shorthand for the multidimensional configuration $\mathbf{x}$). The conformational distribution of the protein without a ligand interaction is given by

$$P(x) = Q^{-1} e^{-G(x)/k_B T}, \quad (14)$$

where $Q$ is the partition function. The partition function $Q$ is related to the total free energy of the protein $G$ as

$$\begin{aligned} G &= \langle G(x) \rangle_x - TS_x \\ &= -k_B T \ln Q \end{aligned}, \quad (15)$$

where $\langle G(x)\rangle_x$ is the mean free energy of individual protein conformations, and $S_x = -k_B \langle P(x)\ln P(x)\rangle_x$ is the entropy of the conformational distribution. Now let $G'(x)$ be the Gibbs free energy of the protein conformation with a ligand interaction, with equations for the conformational distribution $P'(x)$ and free energy $G'$ like those in Eqs. (14) and (15). It follows that $D_x$ is given by

$$D_x = \frac{1}{k_B T}\int dx\, P'(x)\left[-G'(x) + G(x) + k_B T \ln\frac{Q}{Q'}\right], \tag{16}$$
$$= \frac{1}{k_B T}\left\{G' - G - \int dx\, P'(x)[G'(x) - G(x)]\right\}$$

as has been previously mentioned [11]. By the definition of the association constant $K_a(x)$ for the ligand binding to an individual protein conformation, Eq. (16) leads to

$$k_B T D_x = \Delta G + k_B T \int dx\, P'(x)\ln K_a(x), \tag{17}$$

where $\Delta G = G' - G$ is the free energy difference between the protein with and without the ligand interaction. Here and elsewhere, we choose units in which the volume is unity (for other units, $K_a(x)$ should be divided by the volume to use the equations). By the top line of Eq. (15), Eq. (16) also leads to

$$k_B T D_x = -T\Delta S_x + \int dx\, [P'(x) - P(x)]G(x), \tag{18}$$

where $\Delta S_x = S'_x - S_x$ is the conformational entropy difference between the protein with and without the ligand interaction. It is apparent from Equation (18) that $k_B T D_x$ is an allosteric free energy, i.e., the free energy required to change the protein conformational distribution from the equilibrium distribution without the ligand, $P(x)$, to a nonequilibrium distribution $P'(x)$ which is the same as the equilibrium distribution with the ligand bound. This statement about allosteric free energy is essentially the same as Qian's [26] association of relative entropy with the free energy of nonequilibrium fluctuations of a conformational distribution.

## Experimental Measurement of Allosteric Free Energy

How might the allosteric free energy be measured experimentally? Note that Eq. (17) can be rewritten as

$$k_B T D_x = \Delta G + k_B T \int dk_+\, P'(k_+)\ln k_+ - k_B T \int dk_-\, P'(k_-)\ln k_-, \tag{19}$$

where $k_+$ and $k_-$ are the on and off rates for ligand binding. Although the first and third terms of Eq. (19) might be experimentally observable under biologically relevant conditions, the second term involves observing ligand binding to the protein in the ligand-bound conformational distribution, and is not generally observable. However, assuming that the protein may be frozen into a static ligand-bound conformational distribution similar to that at room temperature, analysis methods similar to those used to probe ligand recombination in myoglobin [6] might yield experimental measurements of allosteric free energy, at least in some limited cases. Such a measurement would involve flash-freezing the protein in the ligand-bound form, probing the ligand-binding kinetics at low temperature, and extrapolating the results to high-temperature (Fig. 5).

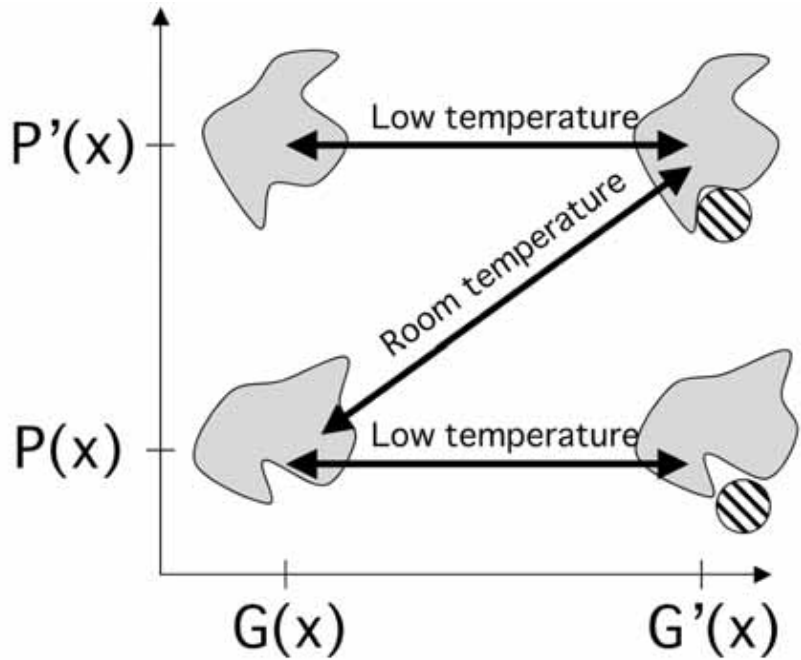

**FIGURE 5.** Experimental observation of ligand-binding kinetics. At room temperature, the protein conformational distribution changes depending upon whether the ligand is bound. At low temperatures, key conformational changes required for relaxation of the conformational distribution are kinetically suppressed, enabling binding kinetics to be observed for a largely "static" protein conformational distribution [6].

## IMPLICATIONS FOR LIGAND BINDING AND ALLOSTERY

### Model Of Allosteric Transitions

Equation (17) may be rewritten as
$$\Delta G = T\Delta D^a - k_B T \int dx \ P'(x)\ln K_a(x) \\ = -T\Delta S_x + \int dx \ [P'(x) - P(x)]G(x) - k_B T \int dx \ P'(x)\ln K_a(x) . \quad (20)$$

In this section the notation is now changed such that the allosteric free energy $k_B T D_x$ is given by $T\Delta D^a$, and, as is commonly done for the entropy, $k_B$ is absorbed into $\Delta D$. The last term in Eq. (20) may be interpreted as the free energy change of the protein upon binding the ligand when the protein conformational distribution is fixed in its final, ligand-bound distribution. The remainder is the free energy required to change the conformational distribution from the initial, ligand-free distribution to the final, ligand-bound distribution. Equation (20) thus describes the total free energy change in a path-dependent process in which the change in the protein conformational distribution occurs independently from ligand binding (Fig. 6). First, the conformational distribution is perturbed from the equilibrium distribution, yielding a (positive) free energy change $T\Delta D^a$. Then, the ligand is bound, yielding an additional free energy change $-k_B T \int dx\, P'(x)\ln K_a(x)$ and a total free energy change $\Delta G$.

Now consider a different path to the same state (Fig. 6): first, the ligand is added to the protein in the ligand-free distribution, changing the equilibrium distribution to $P'(x)$; then, the protein relaxes from the nonequilibrium distribution $P(x)$ to the equilibrium distribution $P'(x)$. Is the free energy change for this path the same as that for the first path? First note that by arguments similar to those leading to Eq. (17), the allosteric free energy $T\Delta D^d$ required to change the protein conformational distribution from an equilibrium ligand-bound form to a nonequilibrium ligand-free form is

$$T\Delta D^d = k_B T \int dx\, P(x)\ln \frac{P(x)}{P'(x)}$$
$$= -\Delta G - k_B T \int dx\, P(x)\ln K_a(x) \qquad (21)$$

which leads to

$$\Delta G = -T\Delta D^d - k_B T \int dx\, P(x)\ln K_a(x)$$
$$= -T\Delta S_x + \int dx\, [P'(x) - P(x)]G'(x) - k_B T \int dx\, P(x)\ln K_a(x), \qquad (22)$$

indicating that the free energies for the different paths are the same. This result is illustrated in the model in Fig. 6.

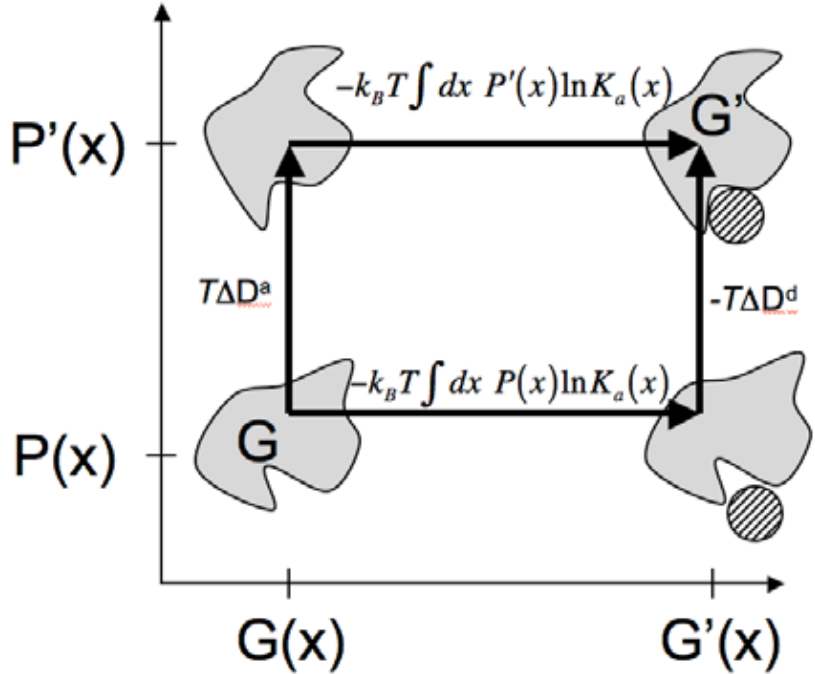

**FIGURE 6.** Two paths for an allosteric transition in the space of probability distributions $P(x)$ and free energy distributions $G(x)$. The protein is depicted using a large shape, and the ligand is depicted using a small circle.

Indeed, the total free energy change in this model is completely independent of the path the system takes in the space of probability distributions $P(x)$ and free energy distributions $G(x)$. For example, assume the following path for ligand binding. (1) The conformational distribution begins at the equilibrium distribution $P(x)$ when the ligand is unbound, and is changed to $\overline{P}(x)$ just before the ligand binds. (2) The free energy distribution is changed discretely from $G(x)$ to $G'(x)$ upon binding the ligand. (3) The conformational distribution relaxes from $\overline{P}(x)$ to the new equilibrium distribution $P'(x)$. By Eq. (18), step (1) entails an allosteric free energy change

$$T\Delta\overline{D} = -T(\overline{S}_x - S_x) + \int dx\, [\overline{P}(x) - P(x)]G(x), \qquad (23)$$

Step (2) then entails a free energy change $-\int \overline{P}(x)\ln K_a(x)$. Finally, step (3) entails an allosteric free energy change

$$-T\Delta\overline{D}' = -T(S'_x - \overline{S}_x) + \int dx\, [P'(x) - \overline{P}(x)]G'(x). \qquad (24)$$

Because of the relation

$$k_B T \int \overline{P}(x)\ln K_a(x) = \int \overline{P}(x)[G(x) - G'(x)], \qquad (25)$$

the total free energy change is the same as that in Eq. (20), indicating that it does not depend on the path (Fig. 7).

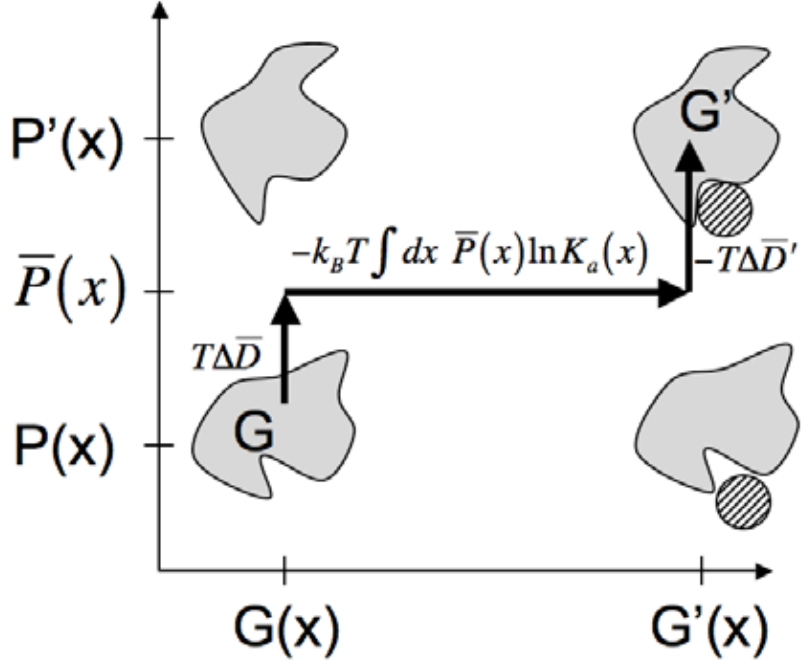

**FIGURE 7.** Abritrary path for allosteric ligand binding. Any paths that share the same beginning and end points in this space lead to the same change in free energy.

## Thermodynamics Of A Molecular Engine

As a practical application of this framework, consider its application to analysis of the thermodynamics of a molecular engine. First, imagine a gedankenexperiment in which a single molecule may be alternately adiabatically stretched and relaxed by an external device, and whose environment may be suddenly changed to alternately cause complete association or dissociation of one or more ligands. This system can be thought of as an idealized real-world system such as titin being stretched using attached beads and laser traps [27]. Figure 8 illustrates a cycle in which the molecule may perform work on the external device. Suppose the molecule begins in a relaxed, ligand-free state with conformational distribution $P(x)$. During the initial stretching stroke, the conformational distribution of the molecule changes from $P(x)$ to $P^\lambda(x)$, which requires a free energy $G^\lambda$-$G$. Now, change the environment to cause ligand binding, leading to a new free energy $G'^\lambda$ and conformational distribution $P'^\lambda(x)$. During the subsequent power stroke, the molecule is allowed to relax, leading to a free energy change $G'$-$G'^\lambda$. Finally, the environment is changed to cause ligand release, returning the system to the initial state.

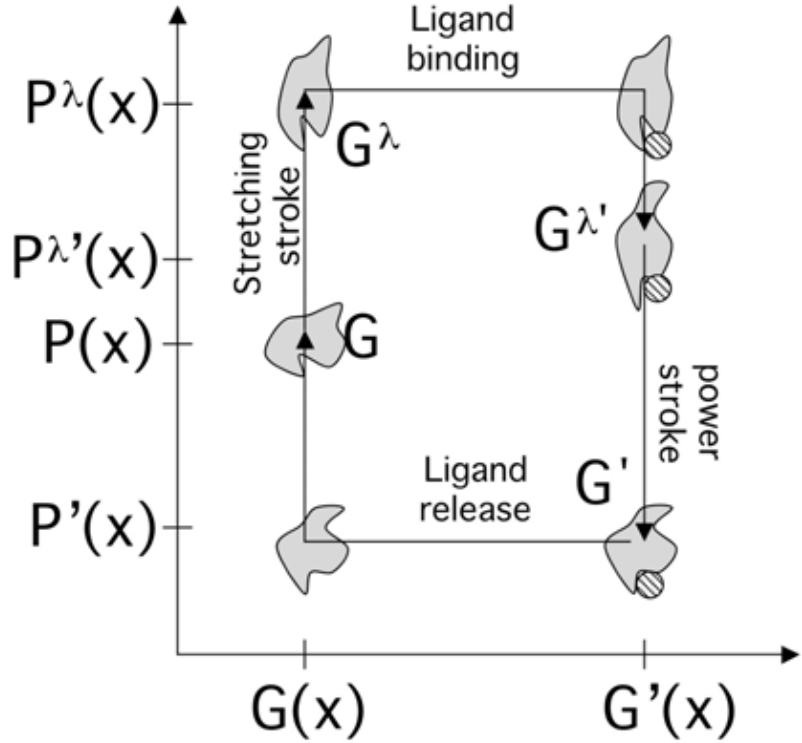

**FIGURE 8.** Modeling thermodynamics of a molecular engine.

By the first law of thermodynamics, the usual expression for the work done on the device is

$$W = G'^\lambda - G' + G - G^\lambda = \Delta G^\lambda - \Delta G. \tag{26}$$

However, using Eq. (18) for the allosteric free energy, the work $W_p$ done by the molecule on the device during the power stroke is

$$W_p = T(S'_x - S'^\lambda_x) - \int dx \, [P'(x) - P'^\lambda(x)] G'(x), \tag{27}$$

and the work $W_s$ done by the device on the molecule during the stretch stroke is

$$W_s = T(S_x - S^\lambda_x) - \int dx \, [P(x) - P^\lambda(x)] G(x). \tag{28}$$

Using Eq. (17), it is easily shown that $W$ as defined in Eq. (26) is equal to $W_p - W_s$. The work may thus be alternatively calculated using binding free energies or, using the present framework, allosteric free energies.

Now consider a couple of ways to make this model more realistic. First, for a real-world system, the concentration of ligand is finite. As an example, let the protein be capable of binding only a single ligand molecule, and let the equilibrium conformational distribution $P_L(x)$ be modeled as a mixture between the ligand-free distribution, $P(x)$, and the ligand-bound distribution, $P'(x)$:

$$P_L(x) = P(x)(1-f) + P'(x)f$$
$$= P(x)\left[(1-f) + K_a(x)f/V\right]. \quad (29)$$

In Eq. (29), $f$ is the fraction of time that a ligand is bound to the protein, and $V$ is the volume of the system. At a free ligand concentration $[L]$,

$$P_L(x) = P(x)\frac{1 + K_a(x)e^{-\Delta G/k_B T}[L]}{1 + Ve^{-\Delta G/k_B T}[L]}, \quad (30)$$

which may be used to interpolate between $G(x)$ and $G'(x)$ in the above equations. Next, it is important to consider that the stretching and power strokes each take a finite time, leading to variability in the work. Assuming Jarzynski's equality [28] holds for this system, as has been experimentally demonstrated for a similar system [29], Eqs. (27) and (28) are then modified as follows

$$k_B T \ln\langle e^{W_p/k_B T}\rangle = T(S'_x - S'^\lambda_x) - \int dx\, [P'(x) - P'^\lambda(x)]G'(x), \quad (31)$$

and

$$k_B T \ln\langle e^{W_s/k_B T}\rangle = T(S_x - S^\lambda_x) - \int dx\, [P(x) - P^\lambda(x)]G(x). \quad (32)$$

## Requirements For Functional Ligands

In light of Eq. (17), it is interesting to revisit the implications of the finding that values of $D_x$ are high for interactions in small-molecule ligand-binding sites of native proteins [11] (Prediction Of Ligand-Binding Sites). On the one hand, this empirical, computational result suggests that $k_B T D_x$ is relatively high for native ligand-binding interactions. On the other hand, because the ligand binds in the binding site, $\Delta G$ should be negative and of relatively high magnitude. In considering the biological requirements of protein-ligand interactions, it is usual to emphasize the importance of ligand affinity in maximizing the total binding energy. However, by rewriting Eq. (17) as

$$k_B T \int dx\, P'(x)\ln K_a(x) = T\Delta D^a - \Delta G, \quad (33)$$

the requirements for the affinities of *functional* ligands are shown to be even more strict than is usually considered. Ligand binding to the protein in the ligand-bound conformational distribution must yield sufficient free energy not only to yield a relatively high binding energy, but also to yield a relatively high allosteric free energy.

Non-negativity of $D_x$ also leads to some interesting relations involving binding affinities. For example, adding Eqs. (17) and (21) leads to the following inequality:

$$\int dx\, [P'(x) - P(x)]\ln K_a(x) > 0. \quad (34)$$

It also follows from Eq. (17) that, if $\Delta G$ is negative, $\int dx \, P'(x) \ln K_a(x)$ is positive, meaning that there is a net loss of free energy when the ligand binds to the protein in the ligand-bound conformational distribution; and from Eq. (21), if $\Delta G$ is positive, $\int dx \, P(x) \ln K_a(x)$ is negative, meaning that there is a net increase in free energy when the ligand binds to the protein in the ligand-free conformational distribution.

## CONCLUSIONS

As reviewed above, previous studies have demonstrated the utility of the Kullback-Leibler divergence between protein conformational distributions before and after ligand binding, $D_x$, in analysis of ligand binding and protein vibrations [9-11]. Two especially significant outcomes of these studies are (1) a prescription for optimizing coarse-grained models of molecular vibrations; and (2) the discovery that native binding-site interactions cause a large change in protein dynamics. The former prescription is a general method that is applicable to coarse-grained modeling of proteins and other materials. The latter discovery enabled the development of a method for predicting ligand-binding sites in proteins [11]. In addition, the latter discovery suggests that sites where interactions cause a large change in protein dynamics might be well-suited to evolve as sites for controlling protein function. It also suggests that ligand-binding sites might tend to evolve at sites that are well-suited for controlling protein function.

In the rest of this paper, the thermodynamic and biophysical significance of $D_x$ was emphasized. The quantity $k_B T D_x$ was shown to be an example of an allosteric free energy: $T\Delta D^a = k_B T D_x$ is the free energy required to change the protein conformational distribution from an equilibrium ligand-free distribution to a nonequilibrium ligand-bound distribution. Equations (17) and (18) enable the allosteric free energy to be calculated and suggest how it might be experimentally measured. They also indicate that allosteric transitions are naturally modeled in terms of motion of the system in the space of conformational distributions $P(x)$ and free-energy distributions $G(x)$.

Interestingly, the present model of allosteric transitions unifies two views of the coupling of ligand binding to conformational changes. In one view, the ligand interaction causes a perturbation in the protein conformational distribution that favors ligand binding; this view is well-aligned with Koshland's early description of induced fit, in which ligand-protein shape complementarity is only achieved when the ligand interacts with the protein [30]. In an alternative view, influence of the ligand on the protein is not significant until a perturbation in the protein conformational distribution leads to an increase in binding affinity (see, e.g., Ref. [31]). In the present context, these two views correspond to the two different pathways in Fig. 6. The first view corresponds to the path that first follows $G(x) \rightarrow G'(x)$, and then follows $P(x) \rightarrow P'(x)$. The second view corresponds to the alternative path that first follows $P(x) \rightarrow P'(x)$, and then follows $G(x) \rightarrow G'(x)$. Both paths lead to the same overall free energy change, and an argument about which of these paths might be preferred for a given system may be resolved by examining the kinetics of the different pathways and determining which is dominant in the context of the model.

As illustrated in Fig. 7, ligand-binding transitions that involve both of the above views are also possible. Determining the most probable path of *all* possibilities therefore requires consideration of not only the extreme paths that correspond to these views, but also of paths such as that illustrated in Fig. 7, and other possibilities not considered here, such as paths that involve continuous changes in the free-energy distribution $G(x)$. For example, the most probable path might first involve a fluctuation in the conformational distribution, then ligand binding, and then a subsequent relaxation to the equilibrium ligand-bound conformational distribution. Such a more general view of the coupling of ligand binding and conformational changes will be necessary to fully understand the nature of allosteric transitions in proteins.

## ACKNOWLEDGMENTS

I am grateful to the Biocomputing and Physics of Complex Systems Research Institute (BIFI) for sponsoring my attendance and lecture at the Second International BIFI Congress in Zaragoza. The research described in this paper was supported by the US Department of Energy.